\documentclass[letterpaper, 10 pt, conference]{ieeeconf}
\IEEEoverridecommandlockouts
\newcommand{\comment}[1]{}

\newcommand{\be}[0]{\begin{equation}}
\newcommand{\ee}[0]{\end{equation}}
\newcommand{\ben}[0]{\begin{equation*}}
\newcommand{\een}[0]{\end{equation*}}
\newcommand{\bena}[0]{\begin{eqnarray*}}
\newcommand{\eena}[0]{\end{eqnarray*}}
\newcommand{\bea}[0]{\begin{eqnarray}}
\newcommand{\eea}[0]{\end{eqnarray}}

\usepackage[english]{babel}
\usepackage{amsthm}
\usepackage[version=4]{mhchem}
\usepackage{siunitx}
\usepackage{array}
\usepackage{longtable}
\usepackage{nomencl}
\usepackage{mathrsfs} 
\usepackage{mathtools} 
\usepackage{optidef}
\usepackage{multicol}
\usepackage{svg}
\usepackage{lipsum}
\usepackage{calrsfs}
\usepackage{scalerel}
\usepackage{stackengine}
\usepackage{booktabs}
\usepackage{caption}
\usepackage{enumerate}
\usepackage{cite}
\usepackage{amssymb,amsfonts}
\usepackage{algorithmic}
\usepackage{graphicx}
\usepackage{textcomp}

\DeclareMathAlphabet{\mathcal}{OMS}{cmsy}{m}{n}
\theoremstyle{definition}
\newtheorem{definition}{Definition}
\newtheorem{theorem}{Theorem}
\newtheorem{lemma}{Lemma}
\newtheorem{assumption}{Assumption}
\newtheorem{remark}{Remark}

\begin{document}

\title{Sensor-Based Safety-Critical Control Using an Incremental Control Barrier Function Formulation via Reduced-Order Approximate Models}

\author{Johannes Autenrieb$^{1}$ and Hyo-Sang Shin$^{2}$
\thanks{$^{1}$ Department of Flight Dynamics and Simulation, Institute of Flight Systems, German Aerospace Center (DLR), 38108 Braunschweig, Germany.
(email: \texttt{johannes.autenrieb@dlr.de})}
\thanks{$^{2}$ 
Cho Chun Shik Graduate School of Mobility, Korea Advanced Institute of Science and Technology (KAIST), Daejeon 34141, Republic of Korea.
(email: \texttt{hyosangshin@kaist.ac.kr}) }
}

\maketitle

\begin{abstract}
The existing control barrier function literature generally relies on precise mathematical models to guarantee system safety, limiting their applicability in scenarios with parametric uncertainties. While incremental control techniques have shown promise in addressing model uncertainties in flight control applications, translating these approaches to safety-critical control presents significant challenges. This paper bridges this gap by introducing measurement-robust incremental control barrier functions (MRICBFs), which leverage sensor-based reduced-order models to provide formal safety guarantees for uncertain systems. By carefully addressing the challenges of sensor accuracy and approximation errors in the incremental formulation, our approach enables substituting specific model components with real-time sensor measurements while maintaining rigorous safety guarantees. This formulation overcomes the limitations of traditional adaptive control methods that adjust system parameters over time, enabling immediate and reliable safety measures for a class of model uncertainties. The efficacy of MRICBFs is demonstrated in two simulation case studies: a simple first-order system with time-varying sensor biases and a more complex overactuated hypersonic glide vehicle with multiple state constraints.

\end{abstract}


\section{Introduction}
\label{SecIntroduction}
The integration of safety-critical control systems is increasingly crucial in autonomous systems, particularly in aerospace. These systems demand high reliability and accuracy due to their complex dynamics and strict safety requirements. Control barrier functions (CBFs) have emerged as a widely embraced safety assurance approach, serving as mathematical constructs that guarantee a system remains within predefined safe regions \cite{ames2016control}. Despite their convenience, traditional CBFs rely on precise mathematical models to represent system dynamics. This dependence on model accuracy can be limiting in the presence of uncertainties, reducing the adaptability and robustness of control strategies \cite{Autenrieb2023}. While several approaches have been proposed to ensure safety for uncertain systems using CBFs \cite{ Jankovic_2018, Buch_2022, Taylor_2020_ml, Ames_2020a, Nguyen_2022}, most employ adaptive control methods. These approaches use adaptive CBFs, where system dynamics are learned online, enabling the controller to compensate for unknown or uncertain behaviors. While this improves adaptability, ensuring strict set invariance during learning remains a challenge, especially in the transient phase where model inaccuracies may compromise safety.\\

Recently, incremental control techniques such as incremental nonlinear dynamic inversion (INDI) and incremental backstepping (IBS) have gained prominence in aerospace applications. These approaches partially substitute model-based information with real-time sensor measurements, enhancing controller robustness to certain parametric uncertainties in flight control \cite{Acquatella2012, Sieberling2010}. This is achieved by reformulating the dynamical model incrementally and replacing erroneous or incomplete non-control input-related model data with sensor-derived quantities. Despite promising empirical results in various practical settings \cite{Lombaerts2019, Autenrieb2023sc}, theoretical stability guarantees for these control approaches were lacking until recently. While the stability of INDI-based concepts has been discussed \cite{Wang2019}, issues such as dependency on sensor accuracy have not yet been fully addressed.\\

Building upon these incremental control strategies, this paper proposes a new notion of sensor-based safety-critical controllers called incremental control barrier functions (ICBFs). The proposed concept extends the benefits of sensor-based control strategies to safety-critical applications. By using real-time sensor measurements to replace specific model components, we can provide guaranteed safety for systems with a particular class of model uncertainties. This approach is designed to increase the controller's resilience against model mismatches without relying on adaptive controllers, which often lack guarantees for forward invariance during the learning phase. To deal with the complexities of the incremental formulation, we introduce measurement-robust incremental control barrier functions (MRICBFs). These functions offer formal safety guarantees even in the worst-case scenarios of errors, specifically those resulting from reduced-order approximations and dependencies on sensor data accuracy.\\

The contributions of this paper are threefold: First, we introduce a new notion of sensor-based control barrier functions, ICBFs, and discuss their benefits in controlling uncertain systems. Second, we propose the MRICBF formulation to address problems in the ICBF approach related to model approximation errors and dependency on sensor accuracy. Third, we examine the proposed formulation through two simulation-based case studies. We compare it with the standard CBF formulation to showcase its effectiveness and capacity to improve the considered systems' robustness and safety.
\section{Preliminaries and Problem Formulation}
\label{SecPreliminaries}
Consider a state space \( \chi \subset \mathbb{R}^n \) and a control input space \( U \subset \mathbb{R}^m \), where it is assumed \( \chi \) is path-connected and \( 0 \in \chi \). Consider the control affine nonlinear system given by:
\begin{equation}
    \label{NonlinearPlant1}
    \dot{x} = f(x) + g(x) u,
\end{equation}
\begin{equation}
    \label{NonlinearPlant2}
    y = p(x)
\end{equation}
where $x \in \chi$, $u \in U$ , $y \in \mathbb{R}^p$, $f: \chi \to \mathbb{R}^n$, $g: \chi \to \mathbb{R}^{n \times m}$, and $p: \chi \to \mathbb{R}^p$
are sufficiently smooth functions. To define safety, we consider a continuously differentiable function  $h: \chi \rightarrow \mathbb{R}$ and a set $S$ defined as the zero-superlevel set of $h$, yielding:
\begin{equation}
    \label{Safe_set_1}
    S \triangleq \begin{Bmatrix} x \in  \chi | h(x) \geq 0 \end{Bmatrix}.
\end{equation}
\begin{definition}[Forward invariance and safety, \cite{Blanchini_1999,Ames_2014}]
The set $S$ is forward  invariant for the system \eqref{NonlinearPlant1}, if for every $x(0) \in S$, $x(t) \in S$ for all $t \in I(x_0) = [0,\tau_{max} = \infty)$. The system \eqref{NonlinearPlant1} is safe with respect to the set $S$ if $S$ is forward invariant.
\end{definition}
\begin{definition}[Class $\mathcal{K}$ functions, \cite{Khalil}]
A continuous function $\alpha: (-b, a) \rightarrow \mathbb{R}$, with $a,b > 0$, is an extended class $\mathcal{K}$ function $(\alpha \in \mathcal{K})$, if $\alpha(0) = 0$ and $\alpha$ is strictly monotonically increasing. If $a, b = \infty$, $\lim_{r \rightarrow \infty} \alpha(r) = \infty$, $\lim_{r \rightarrow -\infty} \alpha(r) = -\infty$ then $\alpha$ is said to be a class $\mathcal{K}_{\infty}$ function $(\alpha \in \mathcal{K}_{\infty})$.
\end{definition}
\noindent
We introduce the notion of a control barrier function (CBF) such that its existence allows the system to be rendered safe w.r.t. $S$ using a control input $u$ \cite{Ames_2014}.
\begin{definition}[CBF,\cite{Ames_2017}]
Let $S \subset \chi$ be the zero-superlevel set of a continuously differentiable function $h: \chi \rightarrow \mathbb{R}$.The function $h$ is a CBF for $S$ for all $x \in S$, if there exists a class $\mathcal{K}_{\infty}$ function $\alpha(h(x))$ such that for the dynamics defined in \eqref{NonlinearPlant1} we obtain:
\begin{equation}
    \label{ControlBarrierFunction_simple}
    \sup_{u\in U} \dot{h}(x,u)  \geq -\alpha(h(x)),
\end{equation}
where
\begin{equation*}
\dot{h}(x,u) = \frac{\partial h}{\partial x}\begin{bmatrix} f(x) + g(x)u \end{bmatrix} =L_{f} h(x) + L_{g} h(x) u
\vspace{0.2cm}
\end{equation*}
\end{definition}

\begin{theorem}
\label{theorem_LCBF}
Given a set $S \subset \chi$, defined via the associated CBF as in \eqref{Safe_set_1}, any Lipschitz continuous controller $k(x) \in K_{S}(x)$  with 
\begin{equation}
    K_{S} (x) = \{ u \in U : L_{f} h(x) + L_{g} h(x) u + \alpha(h(x)) \geq 0 \}
    \label{definition_safe_controller}
\end{equation}
renders the system \eqref{NonlinearPlant1} forward invariant within $S$ \cite{XU2015}.
\end{theorem}

In this paper, we consider a problem where the nonlinear model, described in \eqref{NonlinearPlant1} and \eqref{NonlinearPlant2}, exhibits uncertainties in the unforced system dynamics, given by  
\begin{equation}
\label{ModelProblem1}  
\dot{x} = \hat{f}(x) + g(x)u  
\end{equation}  
\begin{equation}
\label{ModelProblem2}
y = p(x)
\end{equation}
where \( \hat{f}(x) = \Phi f(x) \). It is assumed that \( f(x) \), \( g(x) \) and \( p(x) \) are known, while \( \Phi \in \mathbb{R}^{n \times n} \) is an unknown diagonal matrix with arbitrary entries, expressed as $\Phi = \text{diag}(\phi_1, \phi_2, \dots, \phi_n)$. The objectives are to determine a control input \( u \) for \eqref{NonlinearPlant1} and \eqref{NonlinearPlant2} such that, despite model discrepancies, the system's state $x$ is stabilized and that for any initial condition $x(0) \in S$ the state trajectory satisfies the safety constraint $x \in S$ for all $t \geq 0$. 


\section{Sensor-Based Reduced-Order Approximate Model}
\label{SecSafe_AproxModel}
The output dynamics of the system in \eqref{NonlinearPlant1} and  \eqref{NonlinearPlant2} can be formulated as:
\begin{equation}
    \label{NonlinearPlant4}
    \dot{y} = \frac{\partial p}{\partial x} \dot{x} = \frac{\partial p}{\partial x} f(x) + \frac{\partial p}{\partial x} g(x)u = f(y) + g(y) u.
\end{equation}
A first-order Taylor series expansion-based approximation of \eqref{NonlinearPlant4} around the condition $t - \Delta t$, indicated by the subscript $0$, is defined as:
\begin{equation}
\begin{split}
    \label{Taylor_ICBF}
    \dot{y}= \dot{y}_0 +  A_0 \Delta y + B_0 \Delta u +  \delta(z, \Delta t),
\end{split}
\end{equation}
where
\begin{equation*}
A_0 = \frac{\partial}{\partial y}[f(y) + g(y) u] \Big|_{0}
\end{equation*}
\begin{equation*}
B_0 = \frac{\partial}{\partial u}[f(y) + g(y) u] \Big|_{0}
\end{equation*}
and with $\Delta y \in \mathbb{R}^p$ being the incremental output computed as $\Delta y = (y-y_0)$, $\Delta u \in \mathbb{R}^m$ being the incremental control input defined as $\Delta u = (u-u_0)$ and $z \in \mathbb{R}^p$ being a state somewhere between $y$ and $y_0$. $\delta(z, \Delta t)$ is the remaining error due to truncation, defined as:
\begin{equation}
    \delta(z, \Delta t) = \frac{|f^{k+1}(z)|}{(k+1)!} \Delta y^{k+1}
\end{equation}
with $f = \dot{y}$ and $k$ being the truncation order. In general, the error term is bounded by 
\begin{equation}
    \delta(z, \Delta t) \leq \frac{M}{(k+1)!} |(y-y_0)^{k+1}|,
\end{equation}
where
\begin{equation}
M  = \max_{z\in [y,y_0]}f^{k+1}(z).
\end{equation}
In a generalized way, we can assume that 
\begin{equation}
    |\delta(z, \Delta t)| = \mathcal{O}(\Delta y^2).
\end{equation}
As discussed in \cite{Wang2019}, the norm of the error suggests that for any $\delta > 0$, there exists a $\Delta t > 0$ such that for all $0 < \Delta t < \overline{\Delta t}$,
\begin{equation}
    \|\delta(z, \Delta t)\| \leq \overline{\delta}_e.
\end{equation}
\begin{assumption}
$A_0$ has no eigenvalues with a real part equal to zero.
\label{no_zero_eigenvalues}
\end{assumption}
\begin{lemma}
\label{lemma_staibilty_local}
If the approximate system in \eqref{Taylor_ICBF} is stable in its incremental neighborhood around $y_0$, the system defined in \eqref{NonlinearPlant4} is also stable.
\end{lemma}

\begin{proof}
From \eqref{Taylor_ICBF}, we obtain that $\dot{y} = \dot{y}_0 + A_0 \Delta y + B_0  \Delta u  +\delta(z, \Delta t)$. We can reformulate to obtain  $\dot{y} = A_0 y + B_0  u_c  +\delta(z, \Delta t)$. This leaves us with a formulation that treats $\delta(z, \Delta t)$ as an upper-bounded disturbance on the dynamics. We apply a controller $u_c = K y$ such that $(A_0 + B_0 K)$ is Hurwitz. In that case, it can be seen via the Lyapunov candidate $V(y)= y^T P y$, where $P = P^T >0$ is the solution to $P(A_0 + B_0 K) + (A_0 + B_0 K)^T P =-Q$ with $Q>0$, that \eqref{Taylor_ICBF} has bounded solutions and is stable. By considering Assumption \ref{no_zero_eigenvalues}, the Hartman-Grobman theorem holds \cite{Baratchart2007}, allowing to prove stability for \eqref{NonlinearPlant4}.\\
\end{proof}

For most dynamical systems it is fair to say that the 
influences of incremental changes in the output $\Delta y$ on the dynamics are much smaller than those in the control input $\Delta u$. Examples of such systems can be traditionally found in the aerospace field \cite{Acquatella2012,Lombaerts2019}. Therefore, we neglect the influences in the further control synthesis steps, leading to: 
\begin{equation}
\label{final_first_order_system}
\begin{split}
    \dot{y}=  \dot{y}_0 + B_0 \Delta u  + \sigma (y, \Delta t),
\end{split}
\end{equation}
where the new error function due to the made simplifications is now defined as
\begin{equation}
    \|\sigma(y, \Delta t)\| = \|A_0 \Delta y + \delta(z, \Delta t)\|,
    \label{sigma_error_function}
\end{equation}
due to the continuity of \eqref{NonlinearPlant4} this error is also bounded. We can further state that with $\Delta t \to 0$, the error diminishes
\begin{equation}
    \lim_{\Delta t \to 0} \|\sigma(y, \Delta t)\| = 0.
\end{equation}
This suggests that $\sigma(y, \Delta t)$ becomes negligible for sufficiently high sampling frequencies since $\Delta y \to 0$. It follows consequently that for any $\sigma > 0$, there exists a $\Delta t > 0$ such that for all $0 < \Delta t < \overline{\Delta t}$,
\begin{equation}
    \|\sigma(y, \Delta t)\|_2 \leq \overline{\sigma}_e.
    \label{bound_sigma_e}
\end{equation}
To impose desired output dynamics, we choose the following incremental output controller:
\begin{equation}
\label{incremental_controller}
\Delta u = B_0^{-1}(\nu - \hat{\dot{y}}_0),
\end{equation}
where $\nu = -K \hat{y}$ defines the desired closed-loop dynamics and $\hat{\dot{y}}_0$ a measurement on the time derivative of $y$. We assume that the mapping between the true values and the obtained measurements of $\hat{\dot{y}}_0$ and $\hat{y}_0$ are augmented with additional error functions, such that:
\begin{equation}
    \label{dynamics_measurement_error1}
    \hat{y} = y + e(y)
\end{equation}
and
\begin{equation}
    \label{dynamics_measurement_error2}
    \hat{\dot{y}}_0 = \dot{y}_0 + w(y),
\end{equation}
where  $e : \mathbb{R}^p \to \mathbb{R}^p $ and $w  : \mathbb{R}^p \to \mathbb{R}^p $ are unknown. However, it is possible to formulate bounds on these errors. Without a loss of generality, we can assume that the unknown error functions $e(y)$ and $w(y)$ lie within a compact error sets $E(y)$ and $W(y)$, bounding the measurement uncertainty \cite{Cosner2021}. This leads to the formulation of a maximum error norm, denoted as:
\begin{equation}
\label{bound_measurement_error1}
\epsilon = \max_{e \in E(y)} \|e(y)\|_2
\end{equation} 
and
\begin{equation}
\label{bound_measurement_error2}
\theta = \max_{w \in W(y)} \|w(y)\|_2.
\end{equation} 

\begin{theorem}
If the functions \eqref{sigma_error_function}, \eqref{bound_measurement_error1} and \eqref{bound_measurement_error2} are bounded, such that $\sigma(y, \Delta t)\|_2 \leq \overline{\sigma}_e$, 
$\|e(y)\|_2\leq \epsilon$ and $\|w(y)\|_2\leq \theta$ and the outputs $y$ and $\hat{\dot{y}}_0$ are accessible as measurements, a stable controller, as defined in  \eqref{incremental_controller}, stabilizes the full system defined in \eqref{ModelProblem1} and \eqref{ModelProblem2} with uncertainties.
\end{theorem}

\begin{proof}
The first order approximation for the actual plant is similar to  \eqref{final_first_order_system} defined as $\dot{y}=\dot{y}_0+B_0 \Delta u+\sigma(z, \Delta t)$ with an adjustment on the error term. The function $\sigma(z, \Delta t)$ differs from the one assumed in the controller design since the function contains $\hat{A}_0 \Delta y$, which is the linearization of $\hat{f}(y) = \frac{\partial p}{\partial x} \hat{f}(x)$. Further, $\dot{y}_0$ involves linearized information on $\hat{f}(y)$ since it reflects the offset dynamics of the system. By using the controller defined in \eqref{incremental_controller}, the lack of knowledge on $\hat{f}(x)$ is addressed, leading to the following dynamics once applied:  $\dot{y}=Ky + D$, where $D = K\epsilon - \theta + \overline{\sigma}.$ If $K$ is Hurwitz,  the Lyapunov candidate $V(y)= y^T P y$, with $P = P^T >0$, $Q>0$ and $P K + K^T P =-Q$, shows that even in the case with a deviated $\hat{f}(x)$ term, the system has bounded dynamics, since $\|y\| \leq \sqrt{\frac{2 \lambda_{\max}(P)}{\lambda_{\min}(Q)}} D$, with $\lambda_{\max}$ and $\lambda_{\min}$ being the maximum and minimum eigenvalues of the matrix $Q$ and $P$ \cite{Lavretsky2013_Book}. By considering Lemma \ref{lemma_staibilty_local}, proving the stability of the approximate dynamics proves stability for \eqref{ModelProblem1} and \eqref{ModelProblem2}. 
\end{proof}

\section{Sensor-Based Safety-Critical Control Design}
\label{SecICBF}
With stability guaranteed, we now derive conditions for the safety of the regarded dynamics. We expand the line of reasoning to find guarantees for forward invariance for \eqref{final_first_order_system} and with that for \eqref{NonlinearPlant4}, as long as the error terms are not neglected. We consider the case where the output of the system is a constraint in a zero-superlevel set of $h$ defined as:
\begin{equation}
    \label{Safe_set_2}
    S \triangleq \begin{Bmatrix} y \in  \mathbb{R}^p | h(y) \geq 0 \end{Bmatrix}
\end{equation}
where $h : \mathbb{R}^p \to \mathbb{R}$ is a continuously differentiable function. Even though $S$ is defined in the output space, it can be translated to the state space since the output is a function of the system's state, allowing to formulate state constraints.\\

To increase the readability, the derivation of a safe controller that can provide guarantees in the presence of approximation errors and sensor errors is divided into two steps. First, we will address safety challenges originating from influences of the bounded approximation error term $\sigma (y, \Delta t) $. After that, we will address the combined problem, where the system is rendered safe with additional bounded error terms $e(y)$ and $w(y)$. In the first step, we regard the considered dynamics in \eqref{final_first_order_system}. It is easy to show that, if there exists a function $h$ for the system in \eqref{final_first_order_system}, from here on referred to as incremental control barrier function (ICBF), for all $y \in S$, with a class $\mathcal{K}$ function $\alpha(h(y))=\gamma h(y)$ such that:
\begin{equation}
    \label{Incremental_Control_Barrier_Function}
    \sup_{\Delta u\in \mathbb{R}^m} \dot{h}( y, \Delta u)  \geq -\alpha(h(y)),
\end{equation}
with
\begin{equation*}
\begin{split}
\dot{h}( y, \Delta u) =\frac{\partial h}{\partial x}
[\dot{y}_0 + B_0 \Delta u  + \sigma (y, \Delta t) ],
\vspace{0.2cm}
\end{split}
\end{equation*}
the system stays safe. Implying that any Lipschitz continuous incremental controller $k(y) \in K_{IS}$  with 
\begin{equation}
    K_{IS} (y) = \{ \Delta u \in \mathbb{R}^m : \dot{h}( y, \Delta u)  \geq -\alpha(h(y)) \},
    \label{definition_incremental_safe_controller}
\end{equation}
renders the system \eqref{final_first_order_system} forward invariant within $S$. However, the challenge is that there is no complete knowledge of $\sigma (y, \Delta t)$ that can be used for the control system design. To provide safety guarantees despite the lack of knowledge, we consider the known upper bound, as stated in \eqref{bound_sigma_e}. In analogy to existing concepts of robust control barrier functions (RCBFs), as presented in \cite{JANKOVIC2018}, we introduce the notion of robust incremental control barrier functions (RICBF), in which the existing concepts have been extended and modified to fit the context of the incremental control concepts. From here on, the model information part that is available for the safety-critical controller design is called the nominal function part, defined as $\dot{h}_n(y, \Delta u) = \frac{\partial h}{\partial x}[\hat{\dot{y}}_0 + B_0 \Delta u]$, with $\hat{\dot{y}}_0 = \dot{y}_0$.

\begin{definition}
A continuously differentiable function $h$ is a \textit{robust control incremental barrier function (RICBF)} for \eqref{final_first_order_system} on $S$ if there exists a function $\varphi :  \mathbb{R}^p \times \mathbb{R}_{\geq 0} \rightarrow \mathbb{R}$ and a class $\mathcal{K}$ function $\alpha$ such that the following holds for all $y \in S$:
\begin{equation}
\sup_{\Delta u\in \mathbb{R}^m}  \dot{h}_n(y, \Delta u) - \varphi(y, \Delta t)  > -\alpha(h(y)).
\label{RICBF}
\end{equation}
\end{definition}
The compensation term $\varphi(y, \Delta t)$ is designed to cancel the adverse effects of the bounded approximation error on safety. The existence of such a RICBF implies that any Lipschitz continuous incremental controller $k(y) \in K_{RIS}$  with 
\[
K_{RIS}(y) = \{ \Delta u \in \mathbb{R}^m \mid \dot{h}_n(y, \Delta u) - \varphi(y, \Delta t) \geq -\alpha(h(y)) \}
\]
renders the system \eqref{final_first_order_system} forward invariant within $S$.

\begin{theorem}
\label{theorem_RICBF}
Let $h$ be a RICBF for \eqref{final_first_order_system} on $S$ with $\sigma$ satisfying:
\begin{equation}
\frac{\partial h}{\partial x}[\sigma (y, \Delta t)] + \varphi(y, \Delta t) \geq 0,
\label{RICBF_theorem}
\end{equation}
for all $y \in \partial S$. Then, any controller $\Delta u = k(y) \in K_{RIS}(y)$ renders \eqref{final_first_order_system} safe with respect to $S$.
\end{theorem}
\begin{proof}
It is easy to see by deriving a control input $\Delta u = (\frac{\partial h}{\partial x}B_0)^{-1} (\frac{\partial h}{\partial x}\dot{y}_0-\varphi(y, \Delta t)-\alpha(h(y)))$, from \eqref{RICBF} for which \eqref{RICBF_theorem} holds, that by considering \eqref{Incremental_Control_Barrier_Function} to ensure system safety via $\dot{h}( y, \Delta u)+\alpha(h(y)) \geq 0$ that safety is imposed by the controller, since $\dot{h}( y, \Delta u)\geq 0$.  
\end{proof}

\begin{remark}
In most cases, the most prevalent choice is to set $\varphi(y, \Delta t) = \sup_{y \in \mathbb{R}^p} \|\frac{\partial h}{\partial x} \sigma (y, \Delta t) \|$ \cite{JANKOVIC2018}. Since such a design decision covers the worst-case, the system will be forced to operate within conservative subsets of $S$ when the influences are less significant than assumed. Alternative approaches that may yield less conservative outcomes are discussed in \cite{Alan2023}.
\end{remark}

Now we consider the fact that we are only able to access the corrupted signals of $\hat{\dot{y}}_0$ and $\hat{y}$, as defined in \eqref{dynamics_measurement_error1} and \eqref{dynamics_measurement_error2}. To provide safety guarantees despite the lack of knowledge, we consider the known upper bound, as stated in \eqref{bound_measurement_error1} and \eqref{bound_measurement_error2}. In the following, we extend the introduced concepts of RICBFs and combine them with a concept presented in \cite{Cosner2021}, called measurement-robust control barrier functions, leading to the extension of RICBFs, named measurement-robust incremental control barrier functions (MRICBFs), which are now able to provide safety guarantees for the overall sensor-based reduced-order control concept, in the presence of approximation errors and sensor corruption. We introduce the following Lie notation:
$L_{\dot{y}_0} h(y) = \frac{\partial h}{\partial x}\dot{y}_0 $, $L_{B_0} h(y) = \frac{\partial h}{\partial x}B_0 $, and $L_{\varphi} h(y) = \frac{\partial h}{\partial x}\varphi(y, \Delta t) $.

\begin{theorem}
Given a safe set $S$, defined via the associated CBF as in \eqref{Safe_set_2}. Assume that $L_{\dot{y}_0} h(y)$, $L_{B_0} h(y)$, and $L_{\varphi} h(y)$ are Lipschitz continuous on $S$, with associated Lipschtiz constants $\kappa_{1}$, $\kappa_{2}$, and $\kappa_{3}$ respectively and that $\alpha \circ h$ is Lipschitz continuous on $S$, with an associated Lipschtiz constant $\kappa_{4}$. Consider the functions $w(y)$ and $e(y)$ defined as in \eqref{dynamics_measurement_error1} and \eqref{dynamics_measurement_error2} respectively. We define the functions \( a, b : \mathbb{R}^p \to \mathbb{R} \) as \( a(y) = (\kappa_{1} + \kappa_{3} + \kappa_{4}) e(y) \) and \( b(y) = \kappa_{2} e(y) \). If \( k : \mathbb{R}^p \to \mathbb{R}^m\) is a Lipschitz continuous incremental controller satisfying:
\begin{equation}
\dot{h}_n(y, \Delta u) - \varphi(y, \Delta t)   - \left( a(y) + b(y) \right) \lVert k(y) \rVert_2 \geq -\alpha(h(y))
\end{equation}
for all \( y \in S \), then the system \eqref{final_first_order_system} is safe with respect to \( S \).
\label{theorem_MRICBF}
\end{theorem}

The proof of Theorem \ref{theorem_MRICBF}, which we omit for the sake of brevity, is similar to the proof of Theorem \ref{theorem_RICBF}. We conclude that Theorem \ref{theorem_MRICBF} provides conditions that allow the design of a controller that can render the sensor-based approximate reduced-order model safe while considering approximation errors and sensor errors. The great benefit of the proposed controller is that no accurate knowledge of $f(x)$ is needed anymore during the operation. This allows for substituting missing or wrong information with sensor data as long as error bounds are known.
\section{Results}
\label{SecResults}
\subsection{1D SISO Dynamics}
First, the performance of the proposed MRICBF is compared to a non-robust standard CBF, as suggested in \cite{Ames_2014}. The considered analysis case involves sensors with time-varying biases to the readings, simulating a more challenging operational case with periodic sensor biases modeled as sinusoidal deviations. The simulation employs a simple first-order control system modeled by:
\begin{equation*}
\dot{x} = \Lambda a_p x + b_p u,
\end{equation*}
\begin{equation*}
y = c_p x + e(t),
\end{equation*}
where \(a_p = -1\), \(b_p = 1\) and \(c_p = 1\) define the assumed system dynamics. In the considered case, \(\Lambda\) is set to 0.6 to create a mismatch between the nominal and actual controlled model. The function $e(t)$ being an error term defined as: 
\begin{equation*}
e(t) = \Gamma \sin( \xi t),
\end{equation*}
where \(\Gamma = 0.1\) represents the amplitude of the sinusoidal bias and $\xi$ defines the frequency. The same error is also added to the measurement of \(\hat{\dot{y}}_0 \) and hence $e(t)=w(t)$.  The admissible set of control inputs $u$ is bounded by the maximum limit \(u_{max} = 0.8\) and the minimum limit \(u_{min} = -0.8\). The control task involves tracking a desired, but potentially unsafe, reference signal \(r\) accurately while ensuring the system output \(y\) stays within the safe set $S \triangleq \{ y \in \mathbb{R}^p \mid y_{\min} \leq y \leq y_{\max} \}$. The performance-oriented controller computes a potentially unsafe control input \(\bar{u}\) from the system's state \(y\) and the reference signal \(r\) as follows:
\begin{equation*}
\bar{u} = k_y y + k_r r,
\end{equation*}
where \(k_y\) and \(k_r\) are gain matrices set based on the nominal model, assuming \(\Lambda = 1\), using LQR techniques to minimize the cost function:
\begin{equation*}
J = \int_{0}^{\infty} [x^T Q x + u^T R u] \, dt.
\end{equation*}
Here, \(Q=3\) and \(R=0.2\) are the state and control input weighting matrices chosen to optimally balance state tracking and control effort. The control input \(\bar{u}\) is incrementally adjusted via:
\begin{equation*}
\Delta \bar{u} = \bar{u} - u_{\text{0}},
\end{equation*}
where \(u_{\text{0}}\) is the control input from the previous timestep.\\

We consider the following min-norm QP-MRICBF safety filter:
\begin{argmini*}
    {\Delta u\in \mathbb{R}^m}{\| \Delta u - \Delta \bar{u} \|^2_2}
    {}{}
    \addConstraint{ \dot{h}_n(y, \Delta u) + B_0 \Delta u - \Theta \geq -\alpha(h(x)) }
    \addConstraint{ \Delta u_{min} \leq \Delta u \leq \Delta u_{max} }
\end{argmini*}
where $\Theta = \varphi(y, \Delta t)   + \left( a(y) + b(y) \right) \lVert \Delta u \rVert_2$. The results are displayed in Fig. \ref{fig:time_varying_biases}. It can be seen that the traditional CBF fails to maintain the system within the safe set, illustrating its inability to handle model mismatches and unexpected measurement deviations effectively:
\begin{figure}[h!]
\centering
\includegraphics[width=\columnwidth]{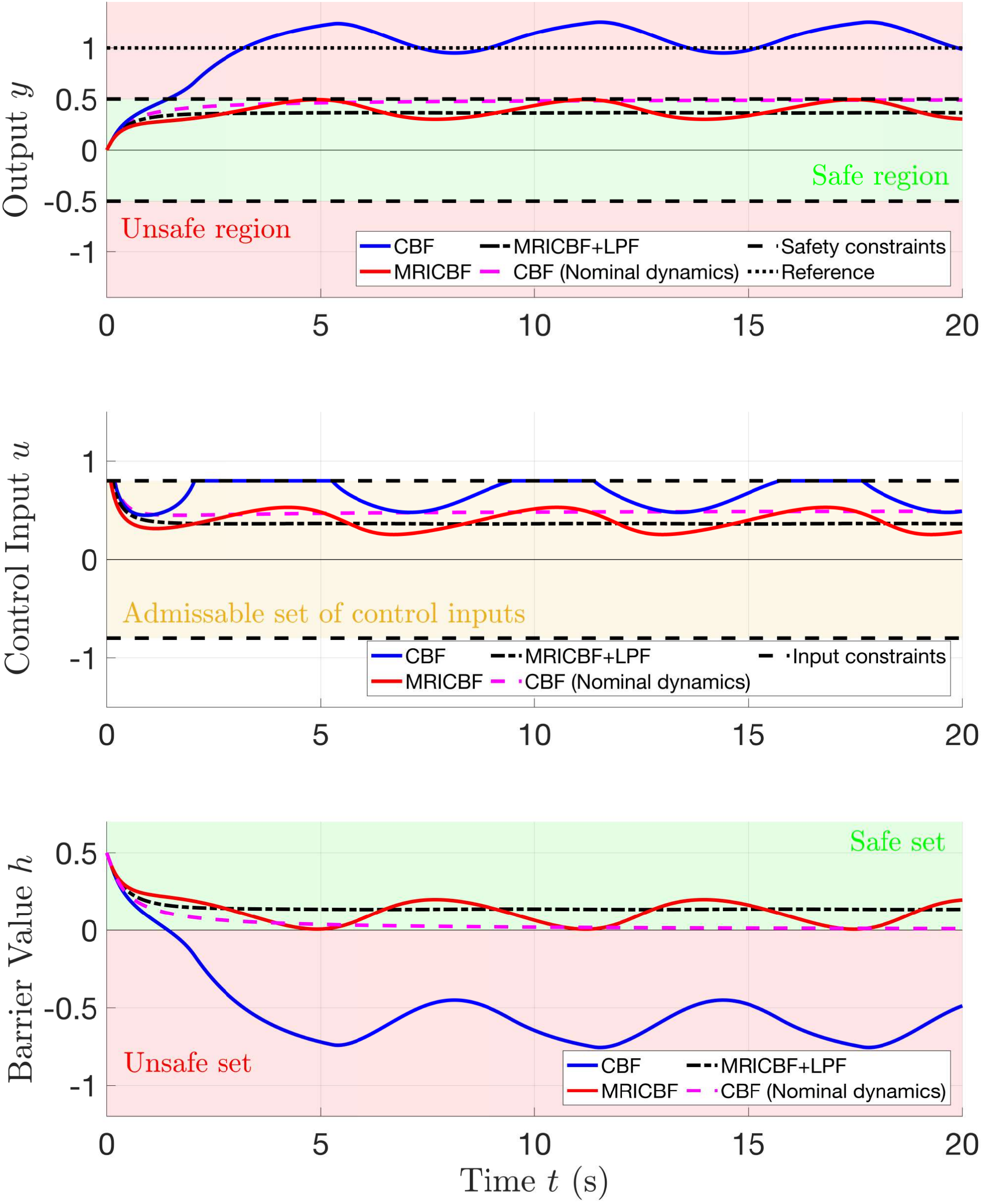}
\caption{System response in the scenario with time-varying sensor biases.}
\label{fig:time_varying_biases}
\end{figure}
In this challenging scenario, the MRICBF demonstrates its strength by effectively accommodating sensor biases and ensuring that the system remains safely within the prescribed limits, thereby highlighting the importance of robust control strategies in unpredictable environments. However, the time-varying bias can still lead to unwanted oscillatory behavior in the MRICBF-based closed-loop system. A low-pass filter (LPF) was applied to the measurement channels to address this, demonstrating that these oscillations can be mitigated. However, it is essential to note that the low-pass filter influences the size of the error set \(E\), which is, to some extent, dependent on the filter's cutoff frequency. While calculating the exact bounds is challenging, it is appropriate to mention this idea as a more practical consideration. As we can see in the results, the LFP addresses the remaining problem and improves the system's safety, leading to a smoother closed-loop response without compromising safety.  
%
\subsection{Overactuated Hypersonic Glide Vehicle}
\begin{figure*}[h!]
\centering
\includegraphics[width=1\textwidth]{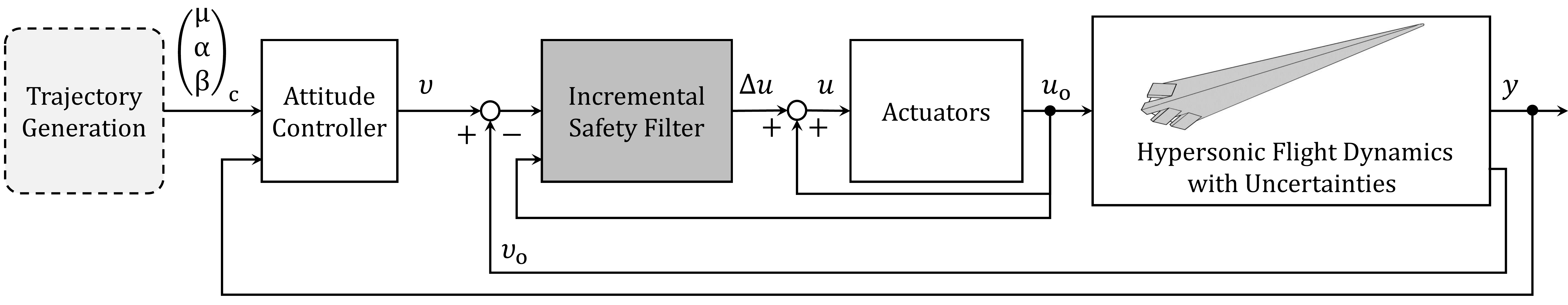}
\caption{Overview of the proposed safety-critical control framework applied to the control of a hypersonic vehicle with parametric uncertainties.}
\label{fig:overview}
\end{figure*}

\begin{figure}
\centering
\includegraphics[width=1\columnwidth]{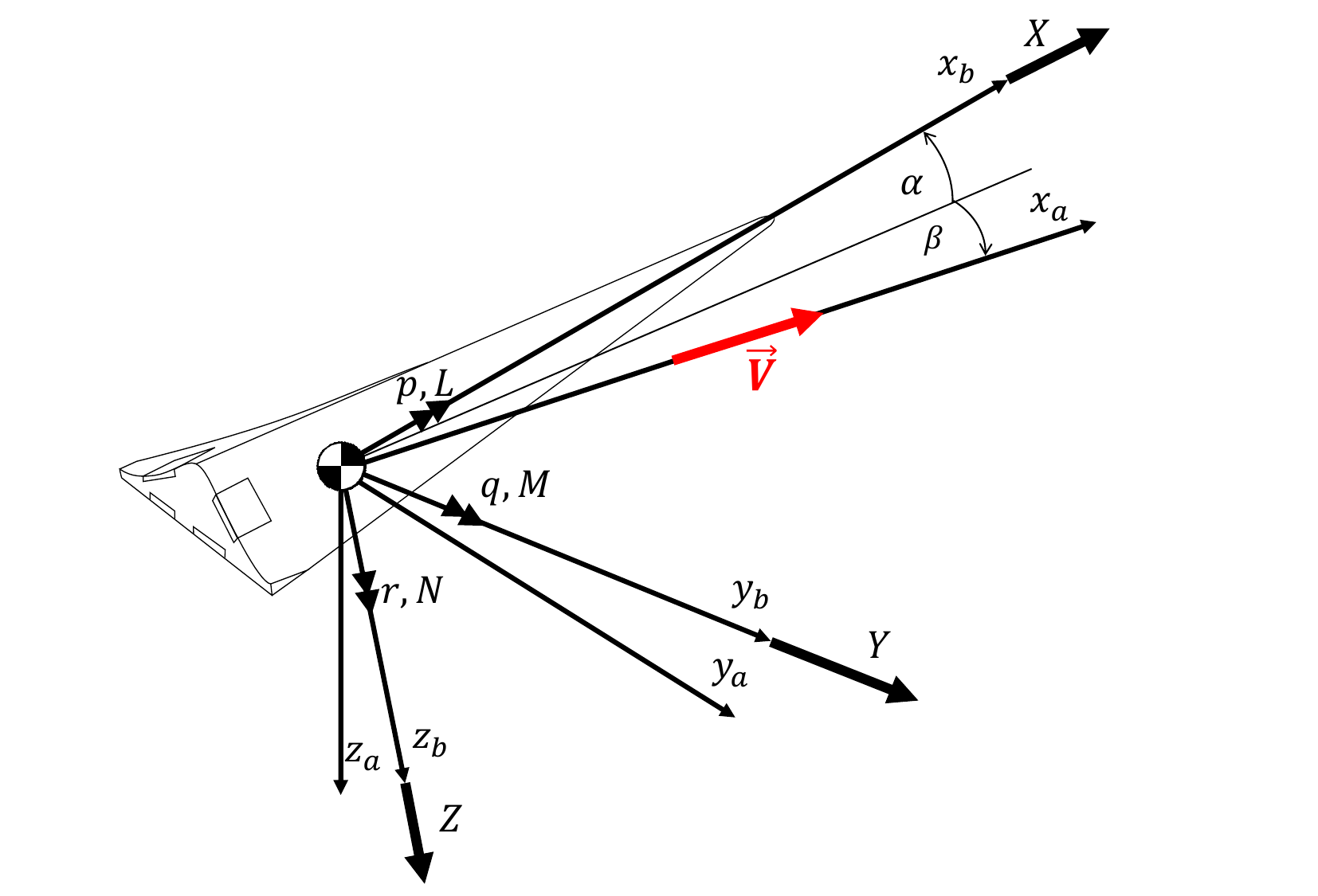}
\caption{Example sketch of external forces and moments acting on a hypersonic glide vehicle~\cite{Autenrieb2021}.}
\label{fig:External forces and moments acting on the GHGV-2 concept}
\end{figure}

\begin{figure}[h]
\centering
\includegraphics[width=1\columnwidth]{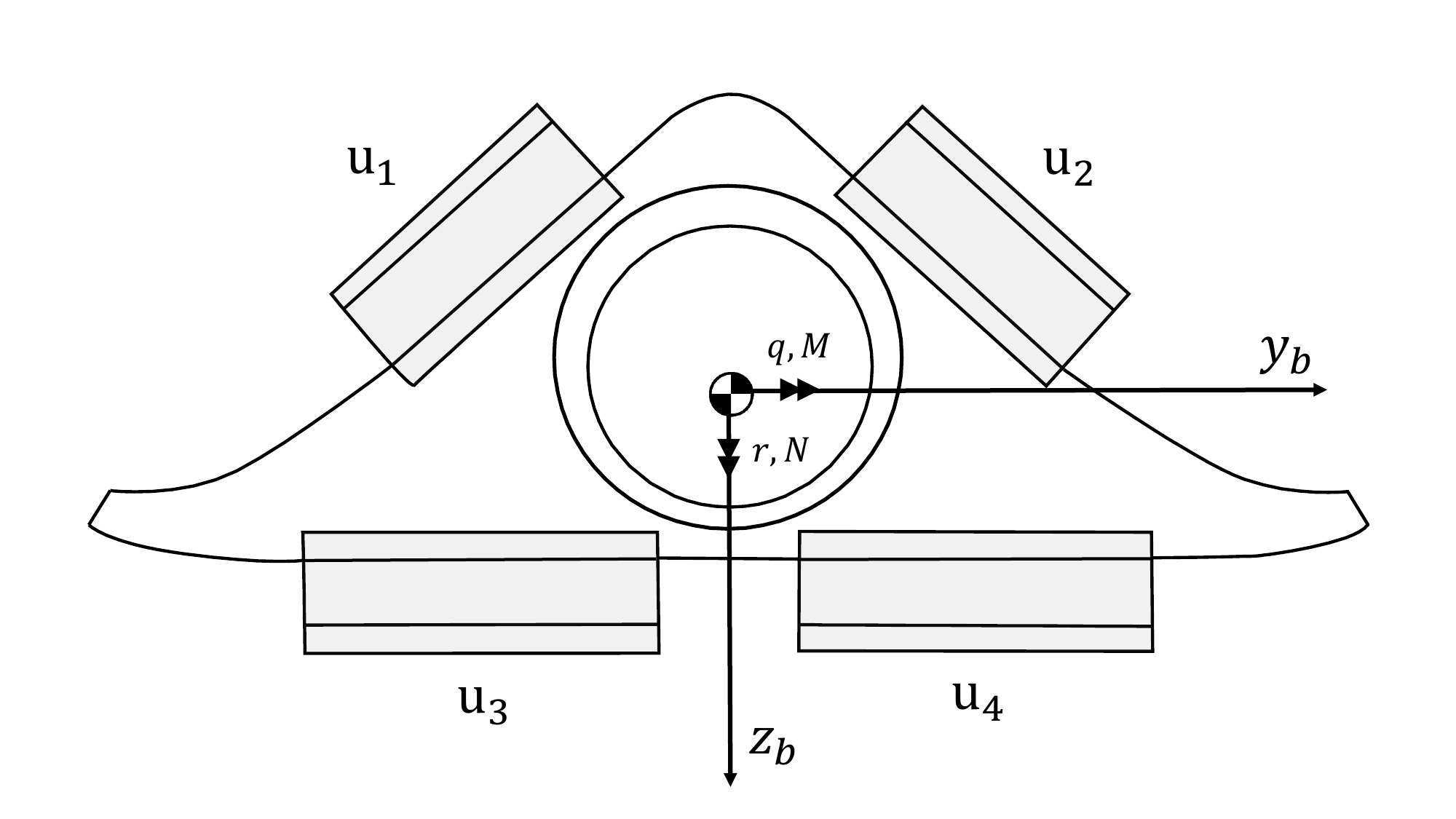}
\caption{Rear view sketch of the considered hypersonic glide vehicle and available control effectors during endoatmospheric operations ~\cite{Autenrieb2022}.}
\label{fig:Available control effectors during endoatmospheric operations}
\end{figure}

This section demonstrates the application of the discussed safety filters to hypersonic glide vehicles through two examples. The concept of the overall control architecture for the considered call of systems is illustrated in Fig. \ref{fig:overview}. 
We begin with a toy example to introduce the fundamental concepts of hypersonic glide vehicles and compare the proposed MRICBF with the standard CBF. We then extend the MRICBF concept to the more complex and realistic case with the DLR GHGV-2 vehicle, incorporating multiple state constraints. Both examples feature an overactuated hypersonic waverider system with four control surfaces. Fig.~\ref{fig:External forces and moments acting on the GHGV-2 concept} displays an example of such a vehicle with the components $X$, $Y$, $Z$ of the total external force vector $\vec{R}$ and the components $L$, $M$, $N$ of the total external moment vector $\vec{Q}$ expressed in the body-fixed frame of the vehicle.
A control problem with simplified dynamics is first considered in this simulation analysis. The relevant isolated pitch rate dynamic of the considered vehicle is defined as:
\begin{align*}
\dot{q} &= \frac{1}{I_{yy}} \left[ \bar{q} S l_{ref} \left( C_{M_0}(\operatorname{Ma}, \alpha) + \frac{C_{M_q}(\operatorname{Ma}, \alpha)}{2V}q \right) + B_p {u}\right]
\end{align*}
where $\alpha$ represents the angle of attack, $q$ the pitch rate and $\operatorname{Ma}$ the Mach number. $\bar{q}$ represents the dynamic pressure (i.e., $\rho V^2/2$), $\rho$ the density of air, $V$ the vehicle velocity, $S$ the aerodynamic reference area, $l_{ref}$ the aerodynamic reference length and $I_{yy}$ the moment of inertia about the y-axis. $C_{M_0}(\operatorname{Ma}, \alpha)$ represents the static aerodynamic pitch coefficient and $C_{M_q}(\operatorname{Ma}, \alpha)$ represents the aerodynamic pitch damping coefficient and $B_p$ being the aerodynamic pitch control effectiveness matrix defined as the following Jacobian:
\begin{equation*}
\label{eqn:Control effectivnes matrix}
B_p = \begin{bmatrix}
\displaystyle\frac{\partial M}{\partial u_1 }&\displaystyle\frac{\partial M}{\partial u_2 }&\displaystyle\frac{\partial M}{\partial u_3 }&\displaystyle\frac{\partial M}{\partial u_4 }
\end{bmatrix},
\end{equation*}
with
\begin{equation*}
\displaystyle {u} = \begin{bmatrix} \displaystyle u_{1} & \displaystyle u_{2} & \displaystyle u_{3} & \displaystyle u_{4} \end{bmatrix}.
\end{equation*}
Fig.~\ref{fig:Available control effectors during endoatmospheric operations} shows the available control surfaces with the connected deflections of the upper left fin $u_{1}$, upper right fin $u_{2}$, lower left fin $u_{3}$ and lower right fin $u_{4}$.
A controller $K(e_q)$ computes the corrective virtual control command $\nu_p$ to eliminate rate control error $e_q=q_c - \widehat{q}$, with  $\widehat{q}$ being the body-rate measurement, and $q_c$ the reference signal. The system's overactuation ($m > n$) leads to a control allocation (CA) problem, as the solution to $\nu_p = B_p {u}$ is not unique.
This simplified example considers two safety constraints on the pitch rate $q$. First $h_1(q) = q_{\text{max}} - q$ and second $h_2(q) = q - q_{\text{min}}$, with $q_{\text{max}} = 10^\circ/\text{s}$ and $q_{\text{min}} = -10^\circ/\text{s}$.
To address both control allocation and safety constraints, we formulate a Quadratic Programming (QP) CA-MRICBF optimization problem:
\begin{argmini*}
{\Delta u\in \mathbb{R}^m,\ \delta}{| \Delta {u} |^2_2 + \phi(\delta)}
{\label{eq:qp_caMRICBF}}{}
\addConstraint{\Delta \nu_p = B_p \Delta {u} + \delta}
\addConstraint{\frac{\partial h_1}{\partial x} \left( \hat{\dot{y}}_0 + B_0 \Delta {u} - \Theta_1 \right) \geq -\alpha(h_1(y))}
\addConstraint{\frac{\partial h_2}{\partial x} \left( \hat{\dot{y}}0 + B_0 \Delta {u} - \Theta_2 \right) \geq -\alpha(h_2(y))}
\addConstraint{\Delta {u}_{\text{min}} \leq \Delta {u} \leq \Delta {u}_{\text{max}} \label{eq:qp_constraint4}}
\end{argmini*}
\begin{figure}[h!]
\centering
 \includegraphics[width=1.01\columnwidth]{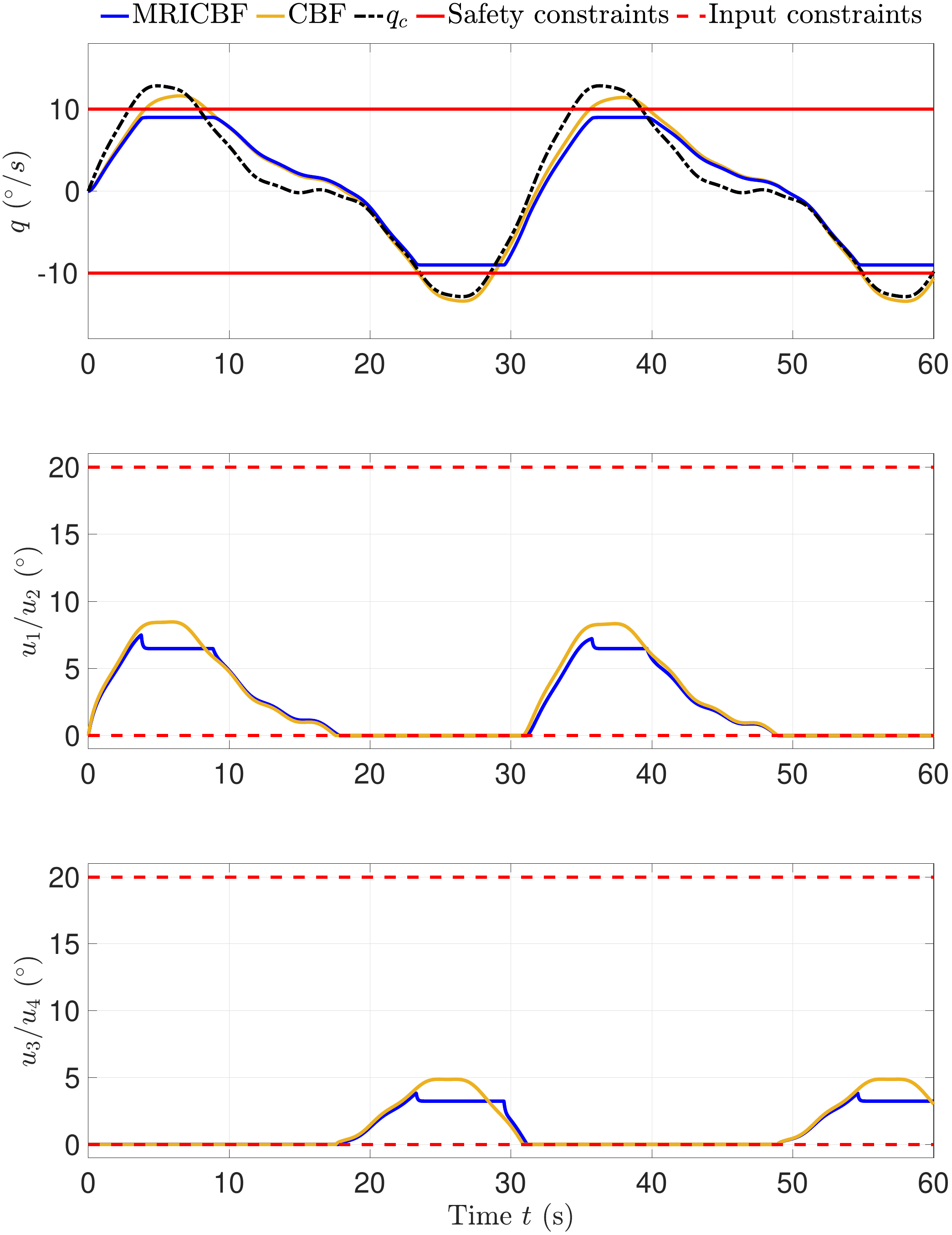}
\caption{Closed-loop response of pitch rate control case with MRICBF for the simplified HGV dynamics.}
\label{fig:MRICBF_HGV}
\end{figure}
In the presented case, the uncertainties have been defined such that there is a $30\%$ mismatch between the assumed aerodynamic parameters $C_{M_0}(\operatorname{Ma}, \alpha)$and $C_{M_q}(\operatorname{Ma}, \alpha)$ and the ones of the plant. Fig. \ref{fig:MRICBF_HGV} displays the discussed pitch rate regulation results for the simplified HGV dynamics case. An unsafe command ($q_c$) is given into the system to analyze the safety filter's ability to ensure safety. The upper graph illustrates the effectiveness of the QP-CA-MRICBF in maintaining pitch rates within safe operational bounds. Conversely, the closed-loop system response with the standard CBF cannot maintain safety and exceeds the set state limits. The time histories of control inputs for the upper flaps $u_1$ and $u_2$, as well as the lower flaps $u_3$ and $u_4$, are provided. Notably, in scenarios requiring pure pitch control, the flaps on both the upper ($u_1$/$u_2$) and lower ($u_3$/$u_4$) sides provide identical inputs. The results affirm the MRICBF's capability to find the right control inputs and maintain safe operational boundaries despite initially unsafe command suggestions.\\

\begin{figure}[h]
\centering
\includegraphics[width=1\columnwidth]{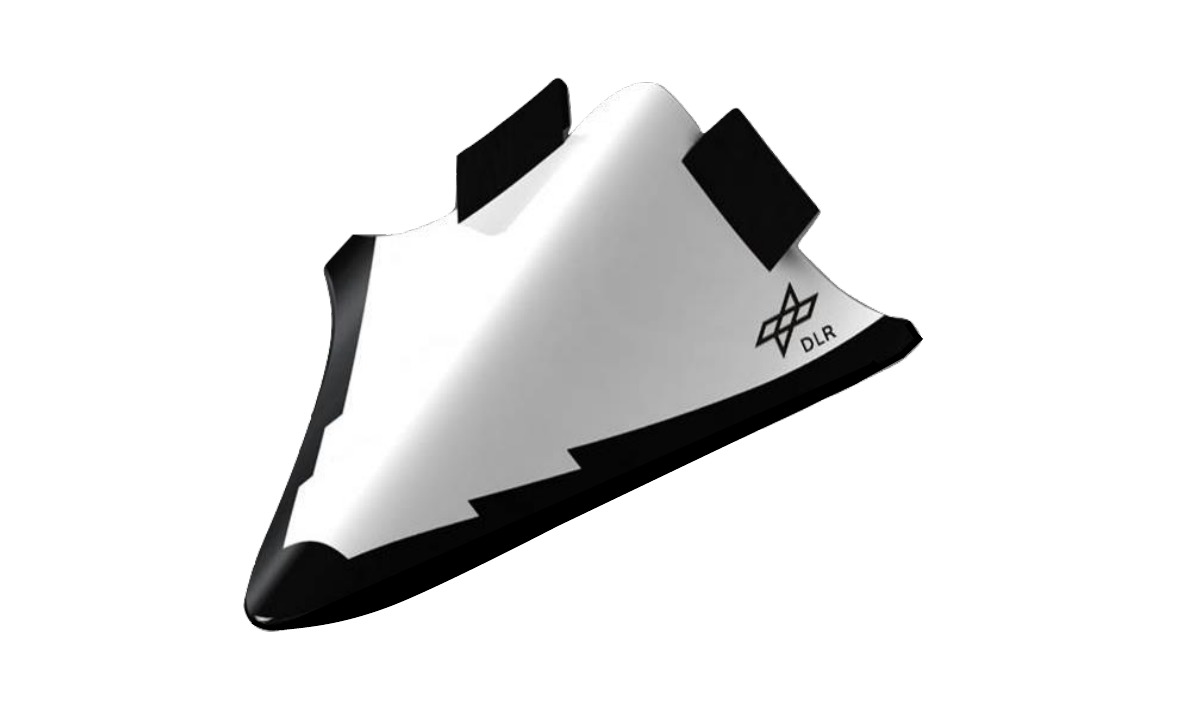}
\caption{The DLR generic hypersonic glide vehicle concept 2 \cite{Gruhn2020}.}
\label{fig:GHGV-Concept}
\end{figure}

\begin{figure}[h!]
\centering
\includegraphics[width=\columnwidth]{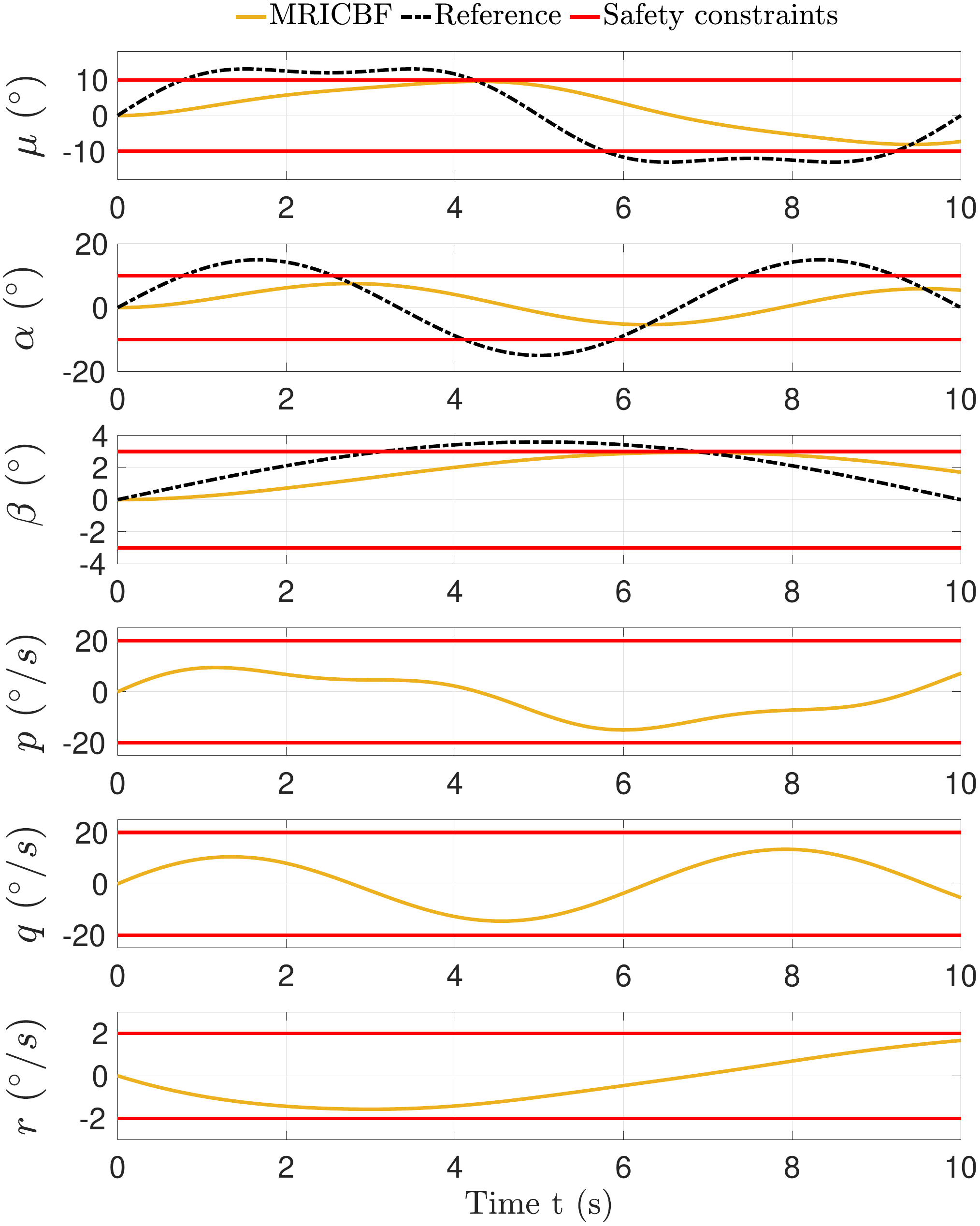}
\caption{Closed-loop response of $\mu$, $\alpha$, $\beta$, $p$, $q$ and $r$ constraints and the system's response trajectory for the GHGV-2.}
\label{fig:state_results_ghgv2}
\end{figure}

\begin{figure}[h!]
\centering
\includegraphics[width=0.9\columnwidth]{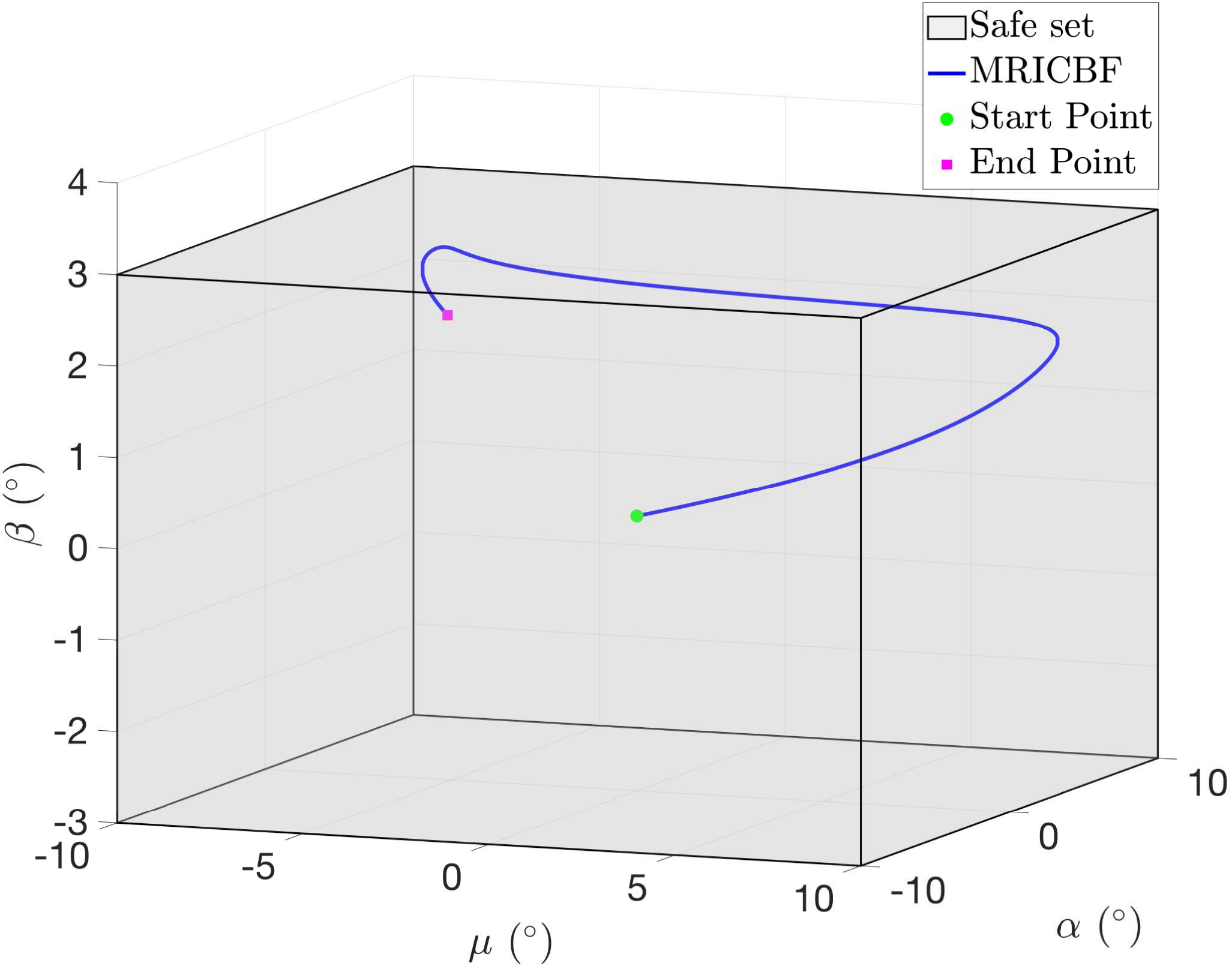}
\caption{3D representation of $\mu$, $\alpha$, and $\beta$ constraints and the system's attitude response trajectory for the GHGV-2.}
\label{fig:state_constraints_ghgv2}
\end{figure}

We now extend these concepts to a more realistic hypersonic waverider configuration developed by a multi-disciplinary development group of DLR that has designed the considered GHGV-2 configuration, shown in Fig. \ref{fig:GHGV-Concept}.
The flight vehicle is based on the waverider concept and is designed to improve lift-to-drag ratios within operations in high Mach number regimes \cite{Gruhn2020}. More information on the dynamics of the system can be found in \cite{Autenrieb2021, Autenrieb2024}. In the regarded case, the dynamics are investigated in which the full attitude dynamics of the system are controlled, leading to a state vector ${x} = \left[ \mu, \alpha, \beta, p, q, r \right]^T$, where $\mu$ is the aerodynamic bank angle, $\alpha$ the angle of attack, $\beta$ sideslip angle and $p$, $q$, and $r$ the angular rates in the body-fixed vehicle frame. The goal is to design a safety filter incorporating constraints on the bank angle $\mu$, angle of attack $\alpha$, and sideslip angle $\beta$. The constraints are $\mu \in [-10^\circ, 10^\circ]$, $\alpha \in [-10^\circ, 10^\circ]$, and $\beta \in [-3^\circ, 3^\circ]$. In the investigated case, sinusoidal reference signal $r$ that intentionally violates the safety constraints is given into the system. Given that the attitude dynamics are second-order, we employ higher-order MRICBFs following the concepts described in \cite{XIAO_2019}. For the full dynamics case of the GHV-2, the uncertainties have been defined such that there is a $20\%$ mismatch between the assumed non-input-related aerodynamic parameters and the ones of the plant. Figure~\ref{fig:state_results_ghgv2} presents the investigation results on the full GHGV-2 model.
The MRICBF response for all states consistently stays within the defined boundaries, while the unsafe command frequently exceeds these limits, showing that the system is rendered safe. Figure~\ref{fig:state_constraints_ghgv2} presents a 3D representation of the constraints on $\mu$, $\alpha$, and $\beta$, along with the system's attitude trajectory. The vehicle must operate within the shaded safe region defined by these state constraints. A blue line represents the system's response trajectory, with a green dot marking the start point and a red dot marking the endpoint.
The trajectory remains entirely within the box, visually confirming the required set invariance.

\section{Conclusions}
\label{secConclusions}
In this paper, we have demonstrated that missing model knowledge for a specific class of model uncertainties can be effectively substituted with real-time sensor information. By formulating controllers that utilize this data, we have addressed the core safety challenges inherent in managing systems with parametric uncertainties. The arising difficulties, particularly those related to sensor accuracy and the robustness of the control system, have been successfully tackled using measurement robust incremental control barrier Functions (MRICBFs). The simulations have confirmed the practical effectiveness of this approach, illustrating that the proposed control strategy maintains system safety and enhances operational reliability under varying conditions. This paper sets a foundational approach for future work to refine the proposed strategies further and extend their application to more complex systems.
\bibliographystyle{IEEEtran}
\bibliography{./Bibliography/references} 

\begin{thebibliography}{10}
\providecommand{\url}[1]{#1}
\csname url@samestyle\endcsname
\providecommand{\newblock}{\relax}
\providecommand{\bibinfo}[2]{#2}
\providecommand{\BIBentrySTDinterwordspacing}{\spaceskip=0pt\relax}
\providecommand{\BIBentryALTinterwordstretchfactor}{4}
\providecommand{\BIBentryALTinterwordspacing}{\spaceskip=\fontdimen2\font plus
\BIBentryALTinterwordstretchfactor\fontdimen3\font minus \fontdimen4\font\relax}
\providecommand{\BIBforeignlanguage}[2]{{%
\expandafter\ifx\csname l@#1\endcsname\relax
\typeout{** WARNING: IEEEtran.bst: No hyphenation pattern has been}%
\typeout{** loaded for the language `#1'. Using the pattern for}%
\typeout{** the default language instead.}%
\else
\language=\csname l@#1\endcsname
\fi
#2}}
\providecommand{\BIBdecl}{\relax}
\BIBdecl

\bibitem{ames2016control}
A.~D. Ames, X.~Xu, J.~W. Grizzle, and P.~Tabuada, ``Control barrier function based quadratic programs for safety critical systems,'' \emph{IEEE Transactions on Automatic Control}, vol.~62, no.~8, pp. 3861--3876, 2016.

\bibitem{Autenrieb2023}
J.~Autenrieb and A.~Annaswamy, ``Safe and stable adaptive control for a class of dynamic systems,'' in \emph{2023 62nd IEEE Conference on Decision and Control (CDC)}, 2023, pp. 5059--5066.

\bibitem{Jankovic_2018}
M.~Jankovic, ``Robust control barrier functions for constrained stabilization of nonlinear systems,'' \emph{Automatica}, vol.~96, pp. 359--367, 2018.

\bibitem{Buch_2022}
J.~Buch, S.-C. Liao, and P.~Seiler, ``Robust control barrier functions with sector-bounded uncertainties,'' \emph{IEEE Control Systems Letters}, vol.~6, pp. 1994--1999, 2022.

\bibitem{Taylor_2020_ml}
A.~Taylor, A.~Singletary, Y.~Yue, and A.~Ames, ``Learning for safety-critical control with control barrier functions,'' in \emph{Proceedings of the 2nd Conference on Learning for Dynamics and Control}, ser. Proceedings of Machine Learning Research, A.~M. Bayen, A.~Jadbabaie, G.~Pappas, P.~A. Parrilo, B.~Recht, C.~Tomlin, and M.~Zeilinger, Eds., vol. 120.\hskip 1em plus 0.5em minus 0.4em\relax PMLR, 10--11 Jun 2020, pp. 708--717.

\bibitem{Ames_2020a}
A.~J. Taylor and A.~D. Ames, ``Adaptive safety with control barrier functions,'' in \emph{2020 American Control Conference (ACC)}, 2020, pp. 1399--1405.

\bibitem{Nguyen_2022}
Q.~Nguyen and K.~Sreenath, ``L1 adaptive control barrier functions for nonlinear underactuated systems,'' in \emph{2022 American Control Conference (ACC)}, 2022, pp. 721--728.

\bibitem{Acquatella2012}
P.~Acquatella, W.~Falkena, E.-J. van Kampen, and Q.~P. Chu, ``Robust nonlinear spacecraft attitude control using incremental nonlinear dynamic inversion.'' in \emph{AIAA Guidance, Navigation, and Control Conference}, Minneapolis, Minnesota, US, 5 2012.

\bibitem{Sieberling2010}
S.~Sieberling, Q.~P. Chu, and J.~A. Mulder, ``Robust flight control using incremental nonlinear dynamic inversion and angular acceleration prediction,'' \emph{Journal of Guidance, Control, and Dynamics}, vol.~33, no.~6, pp. 1732--1742, 2010.

\bibitem{Lombaerts2019}
T.~Lombaerts, J.~Kaneshige, S.~Schuet, G.~Hardy, B.~L. Aponso, and K.~H. Shish, ``Nonlinear dynamic inversion based attitude control for a hovering quad tiltrotor evtol vehicle,'' in \emph{AIAA Scitech 2019 Forum}, January 2019.

\bibitem{Autenrieb2023sc}
J.~Autenrieb, ``Data fusion-based incremental nonlinear model following control design for a hypersonic waverider configuration,'' in \emph{AIAA SciTech Forum 2023}, 2023.

\bibitem{Wang2019}
X.~Wang, E.-J. van Kampen, Q.~Chu, and P.~Lu, ``Stability analysis for incremental nonlinear dynamic inversion control,'' \emph{Journal of Guidance, Control, and Dynamics}, vol.~42, no.~5, pp. 1116--1129, 2019.

\bibitem{Blanchini_1999}
F.~Blanchini, ``Set invariance in control,'' \emph{Automatica}, vol.~35, no.~11, pp. 1747--1767, 1999.

\bibitem{Ames_2014}
A.~D. Ames, J.~W. Grizzle, and P.~Tabuada, ``Control barrier function based quadratic programs with application to adaptive cruise control,'' in \emph{53rd IEEE Conference on Decision and Control}, 2014, pp. 6271--6278.

\bibitem{Khalil}
H.~K. Khalil, \emph{{Nonlinear systems; 3rd ed.}}\hskip 1em plus 0.5em minus 0.4em\relax Upper Saddle River, NJ: Prentice-Hall, 2002.

\bibitem{Ames_2017}
A.~D. Ames, X.~Xu, J.~W. Grizzle, and P.~Tabuada, ``Control barrier function based quadratic programs for safety critical systems,'' \emph{IEEE Transactions on Automatic Control}, vol.~62, no.~8, pp. 3861--3876, 2017.

\bibitem{XU2015}
X.~Xu, P.~Tabuada, J.~W. Grizzle, and A.~D. Ames, ``Robustness of control barrier functions for safety critical control,'' \emph{IFAC-PapersOnLine}, vol.~48, no.~27, pp. 54--61, 2015, analysis and Design of Hybrid Systems ADHS.

\bibitem{Baratchart2007}
L.~Baratchart, M.~Chyba, and J.-B. Pomet, ``A grobman-hartman theorem for control systems,'' \emph{Journal of Dynamics and Differential Equations}, vol.~19, 03 2007.

\bibitem{Cosner2021}
R.~K. Cosner, A.~W. Singletary, A.~J. Taylor, T.~G. Molnar, K.~L. Bouman, and A.~D. Ames, ``Measurement-robust control barrier functions: Certainty in safety with uncertainty in state,'' in \emph{2021 IEEE/RSJ International Conference on Intelligent Robots and Systems (IROS)}, 2021, pp. 6286--6291.

\bibitem{Lavretsky2013_Book}
E.~Lavretsky and K.~A. Wise, \emph{Robust Adaptive Control}.\hskip 1em plus 0.5em minus 0.4em\relax London, UK: Springer London, 2013.

\bibitem{JANKOVIC2018}
M.~Jankovic, ``Robust control barrier functions for constrained stabilization of nonlinear systems,'' \emph{Automatica}, vol.~96, pp. 359--367, 2018.

\bibitem{Alan2023}
A.~Alan, T.~Molnár, A.~Ames, and G.~Orosz, ``Parameterized barrier functions to guarantee safety under uncertainty,'' \emph{IEEE Control Systems Letters}, vol.~PP, pp. 1--1, 01 2023.

\bibitem{Autenrieb2021}
J.~Autenrieb, N.~Fezans, P.~Gruhn, and J.~Klevanski, ``Towards a {C}ontrol-{C}entric {M}odelling and {S}imulation-{F}ramework for {H}ypersonic {G}lide {V}ehicles,'' in \emph{German Aeronautics and Space Congress (DLRK)}, Sep. 2021.

\bibitem{Autenrieb2022}
{ J. Autenrieb and N. Fezans}, ``{N}onlinear {M}odel {F}ollowing {C}ontrol {D}esign for a {H}ypersonic {W}averider {C}onfiguration,'' in \emph{{CEAS} {EuroGNC} "{C}onference on {G}uidance, {N}avigation and {C}ontrol"}, 5 2022.

\bibitem{Gruhn2020}
P.~Gruhn, ``{D}esign and {A}nalysis of a {H}ypersonic {G}lide {V}ehicle ({O}riginal {G}erman {T}itle: {A}uslegung und {A}nalyse eines hypersonischen {G}leitflugk\"{o}rpers),'' in \emph{Conference on Applied Research for Defense and Security in Germany}, Mar. 2020.

\bibitem{Autenrieb2024}
J.~Autenrieb and N.~Fezans, ``\BIBforeignlanguage{en}{Flight control design for a hypersonic waverider configuration: A non-linear model following control approach},'' \emph{\BIBforeignlanguage{en}{CEAS Space Journal}}, pp. 1--24, Apr. 2024.

\bibitem{XIAO_2019}
W.~Xiao and C.~Belta, ``Control barrier functions for systems with high relative degree,'' in \emph{2019 IEEE 58th Conference on Decision and Control (CDC)}, 2019, pp. 474--479.

\end{thebibliography}

\end{document}